# An Analysis of the Recovery Path of the Consumer Sector in the Post-Pandemic Era

Wenbo Lyu*, Jiayi Zhu, Yunan Ding, Keming Zhang

Saxo Fintech Business School, University of Sanya, Sanya, China
Email: *wenbolv@sanyau.edu.cn

## Abstract

This paper proposes a referencable pattern of the recovery of the consumption sector, a new dimension to observe and evaluate the intrinsic value of the consumption sector, and proposes the concept of *sensory-based consumption* and the ranking of the weights of different categories; creates the concept of *digital consumption index*, coupled with *digital RMB index* and *China-style digital economy index*. Finally, we explain the internal logic of digital consumption as a consumption upgrade tool and a higher valuation target in the context of China's economic performance in 2022 and the Chinese government's policy in 2023, leading to the investment strategy of roller conduction effect.

## 1. Introduction

Based on the development experience of countries and regions where the epidemic was released earlier, this article pointes out a recovery path of the consumer sector in the post-pandemic era from Fund Managers' Vision, hoping to help China's economy achieve a faster recovery and avoid potential risks.

It defines somatosensory consumption and refines the valuation rules for consumer companies. It explains the internal logic of digital consumption as a consumption upgrade tool and a higher valuation target in the context of China's economic performance in 2022 and the Chinese government's policy in 2023, leading to the investment strategy of roller conduction effect.

From the perspective of investors, it affirmed the contribution of the digital economy, raised higher expectations for the digital renminbi. This article is lack of econometric models and empirical research on the drum effect, and the author's follow-up research will deepen this.

The article is organized as follows:

It first displays the concept of somatosensory consumption, points out the differences in the valuation of different types of consumption, then explains the importance logic of China's consumption recovery, comments and predicts the policies to stimulate the economy, and finally summarizes the views from the perspective of investors, and puts forward the roller effects and countermeasures.

## 2. Definition

According to the experience of Singapore, Hong Kong and other places after the release of pandemic policy, the conduction and rhythm of the recovery of the consumer sector shows a certain temporal and spatial pattern, generally from high consumption to low consumption, cross-border consumption to local consumption, current consumption to future consumption. Among them, high-end hotel and tourism consumption rose earlier, followed by high-end food and wine consumption, again cross-border service consumption, gradually turning to local consumption sectors, such as the restaurant industry, film industry, etc. (Prasad, 2021).

Normally, the fund manager's investment sequence will be able to follow this approach to arrange the movement of funds, and the industry routine research are also carried out from this spatio-temporal logic. However, this alone from the perspective of spatio-temporal experience, will ignore many different segments of the industry sector valuation differences, combined with the A-share and Hong Kong stock markets in the same sector of different kinds of stock price trends, this paper research team proposed to a more refined classification dimension, that is, the concept of sensory- and gesture-based consumption classification.

People's utility of products and services in consumption can be classified according to sensory, because as human beings with various sensory tubes, all utility will be implemented to the satisfaction of one or several senses. The marginal utility of the five human senses of sight, sound, smell, touch and taste is different, which also provides a logical basis for the focus of different sensory consumption. That is, the valuation rules of different sensory consumption of the sector will be different, in the high valuation sector is more likely to share price anomalies. In the field of investment research, it is necessary to understand and master this valuation habit to facilitate more informed decisions by investors.

According to the dimension of sensory-based consumption is divided into six categories (purely based on the authors' own observations), which are:

1) Tasting: Fast-moving consumer goods (FMCG) products of beverage and alcoholic beverages, necessary food and spices, food for special populations (milk powder, dairy products and nutritional products, etc.), essential medical and pharmaceutical products, etc.

2) Seeing: film and television dramas, events and tournaments, film and television equipment, ARVR products, etc.

3) Hearing: music, concerts, headphones, speaker equipment, etc.

4) Touching: durable goods, household equipment, massage products, some FMCG products, must-type clothing, birth control products, toys and games products, etc.

5) Smelling: perfume, luxury goods, some FMCG products, etc.

6) Mixture Sense: digital consumption (related to digital products or digital channels such as cell phones and computers and other comprehensive digital category), travel, home purchase and rental (and other accommodation needs category), decoration and furnishing category, transportation category, high-end functional clothing, clothing and luggage, medical beauty and make-up health category, health care products and health care equipment, professional sports equipment, education and training category, etc.

According to the experience of valuation weights and stock price anomalies, from large to small should be:

$$\text{Seeing} > \text{Hearing} > \text{Smelling} > \text{Touching} > \text{Tasting}$$

Comprehensive service sense depends on the specific combination of which senses to superimpose the weighting sort. For example, cell phone digital products, superimposed on the sense of sight, hearing, touch multi-sensory, so the valuation weight will be much larger, but also more likely to produce valuation anomalies.

The specific quantification of weights has not been fully conclusive, but it has to be said that the results of such quantification usually change dynamically from year to year, and we can currently achieve qualitative and categorical studies of sector and stock valuations in order of weighting.

Through the study of consumption in developed countries, we found that the influence of digital consumption is huge in the overall consumption sector. This research team recommends a separate index for digital consumption.

$$\begin{aligned}
&\text{Digital Consumption Index} \\
&= \{\text{total digital consumption}\,[\text{number of digital products purchased per capita} \\
&\quad (\text{weighted by}: \text{Seeing} > \text{Hearing} > \text{Smelling} > \text{Touching} > \text{Tasting}] \\
&\quad \times \text{number of daily active per capita} \times \text{active time per capita} \\
&\quad \times \text{total number of active people}\} / \ln \text{total active time}
\end{aligned}$$

$$\text{Degree of Digital Consumption} = \frac{\text{total digital consumption}}{\text{total consumption}}$$

According to our team's previous research on digital RMB index and China-style digital economy index (Su et al., 2022), it is suggested that the ratio D1 (digital consumption index/digital economy index) and ratio D2 (digital RMB in-dex/digital consumption index), both of which can be coupled to assess the weight and stage of digital consumption in promoting the development of digital economy (Ding et al., 2022), and also to quantify (Meaning et al., 2018) the effect of digital RMB development on promoting digital consumption (Minesso et al., 2020).

## 3. Logics

Through the above concepts, we explore the logic of consumer recovery.

We find that Consumer Confidence Index (CCI) in China plummeted in 2022 (see Figure 1), while deposits increased by 24.26 trillion. We briefly divide consumer groups into three major categories: the higher class (with a small proportion), the middle class and the new middle class (with a gradually increasing proportion, but less than 300 million people overall), and the lower class (with a decreasing proportion year by year, but still the largest proportion group). Its higher production group, the rich group, is mainly composed of business people; the middle class and new middle class are mainly intellectuals, white-collar people and other middle and high position groups, as well as civil servants and career staff; the lower class is mainly composed of the vast number of farmers and people without stable occupation and no basic income.

According to McKinsey, during the pandemic, the marginal spending desire of low-income earners in overall consumption decreased, while the total consumption of FMCG products of middle and high-income earners increased rather than decreased (focusing on beverages and makeup and beauty, etc.); the proportion of stockpiling behavior of low-income earners increased, and e-commerce platforms, especially cheap goods platforms, benefited more. Low-income earners are more inclined to save, and during the pandemic period when the overall proportion of social investment and financing is reduced, the counter-intuitive increase in savings undoubtedly increases the overall operating costs of the social and financial system (e.g., increased difficulty in financing the national debt and local debt and increased pressure on debt servicing), which is not conducive to economic recovery.

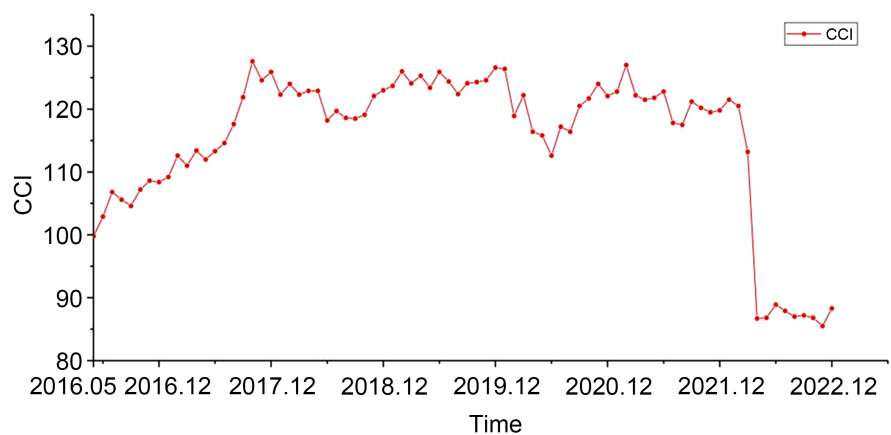

Figure 1. Consumer Confidence Index (CCI) (created by authors).

Before the pandemic policy was released in 2022, China's monetary base M2 continued to break new highs in recent years, with the M2 balance at the end of November at 264.7 trillion yuan, up 12.4% year-on-year, breaking the highest value since April 2016 (see Figure 2). It is very worthwhile to explain the peculiarity of the benchmark comparison year of 2016, because the impact of the sharp pullback in the A-share stock market in 2015, China's monetary and fiscal policies were very accommodative in 2016, and as a result, the whole society experienced a round of house price and food price increases that have been relatively rare since the reform and opening up, so the breakthrough of M2 data at the end of 2022 to a high since April 2016 is an event that deserves high attention. In contrast, not only did social finance and social consumption not rise in the same proportion during the same period, but both even declined, and medium- and long-term loans rose to a limited extent, well below the expectations of the top designers (Kuckertz et al., 2020). These macro indicators have shown that although China's monetary policy has continued to ease in order to stimulate the economy through the trough of the pandemic (Agur et al., 2019), the newly added money has clearly occurred in the financial sector idle phenomenon (Boateng et al., 2022), this monetary trap, if not broken in time, will only increase the overall social CPI, intensify the gap between the rich and poor, further damage the real economy, and fall into a vicious circle (Prasad, 2021). So our government in 2022 with unprecedented strength and tools to curb the rate of CPI rise, the latest CPI data is 1.6% (see Figure 3) (the CPI of major countries in Europe and the United States in the same period are above 5% (Hall et al., 2023), and even some months touched 10%) (Calinescu et al., 2023), the most typical is the price of pork occurred in a roller coaster of ups and downs, the pork cycle and debt cycle have occurred more obvious deformation leveling. These joint efforts of the official and society as a whole are aimed at providing the necessary prerequisites and pavement for a more accommodative monetary and fiscal policy in 2023 (Cai & Zhang, 2022). The latest announcement of China's GDP growth of 3% for the year and 2.9% for the fourth quarter of 2022 (see Figure 4), although somewhat different from the GDP growth rate intuitively felt by the public (subjective feelings are often easily amplified in the darkest periods of the economy) (Castro et al., 2022), is a base figure is quite conducive to the implementation of a economic strategy to use the recovery in consumption to boost GDP in 2023 (Langdana, 2022).

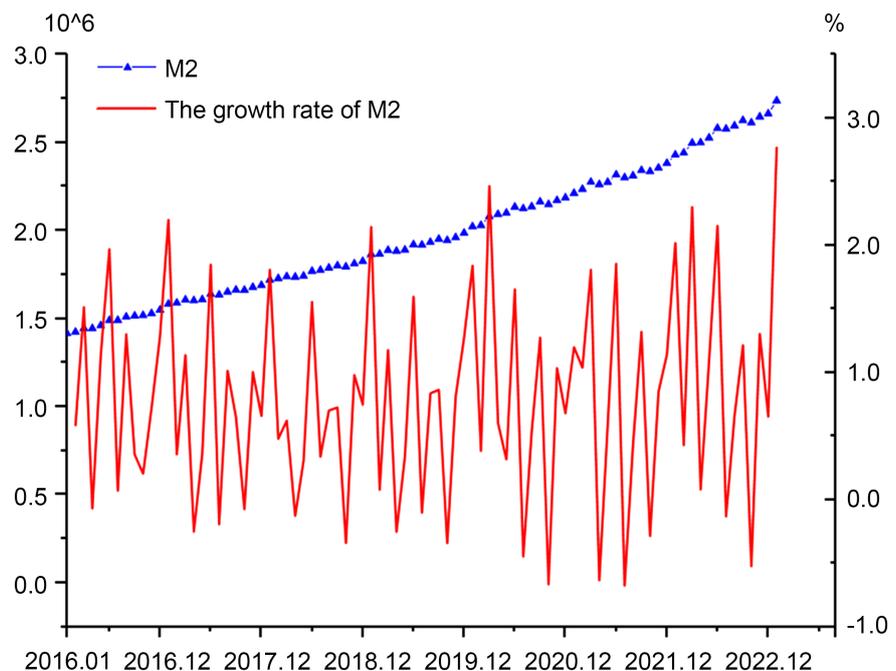

Figure 2. The growth rate of M2 (created by authors).

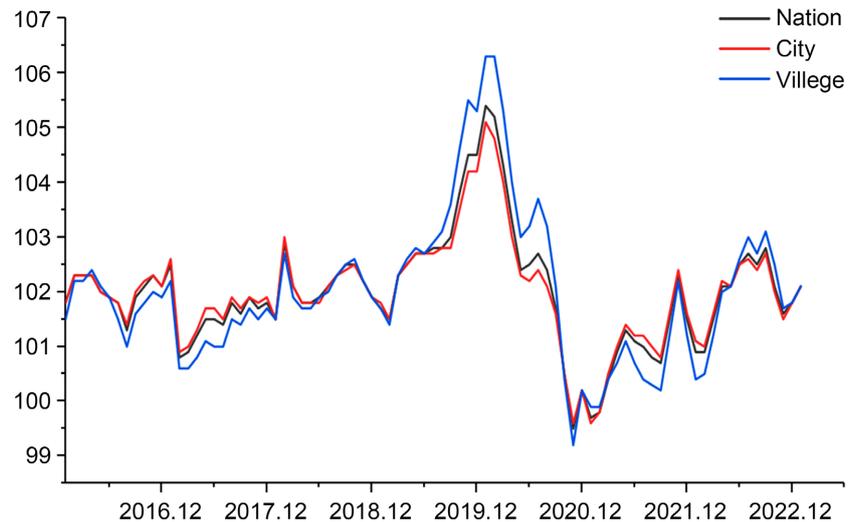

Figure 3. China's CPI. (created by authors).

Our government's national GDP target for 2023 is around 5%, with Hainan Province even proposing a target of 9.5%. In anticipation of a generally bearish overseas market and declining net export indicators in 2023 (Son et al., 2022), if we want to achieve such a high target, the government should not only make further efforts in macro policies to promote consumption (Liu & Liu, 2022), but also make more innovations in actual consumption scenarios and consumer groups and even sources of consumption funds (Su et al., 2022), so that macro policies can be fully implemented and overall operating costs of society can be reduced, and the supply and demand sides can be synchronized to achieve greater (Lyu & Chen, 2023). It can be said that 2023 is a promising year to test the governing ability and economic development ability of local governments (both in southeastern coastal provinces and inland provinces) (Nicola et al., 2020).

There are more and more signs that in 2023, the Chinese government may make big moves to eliminate the disparity between the rich and the poor in specific areas, in other words, in the new year, the high-income group may have to pay more social responsibility, while the middle and low-income group will be guided and supported in the consumption upgrade (Pieterse, 2002).

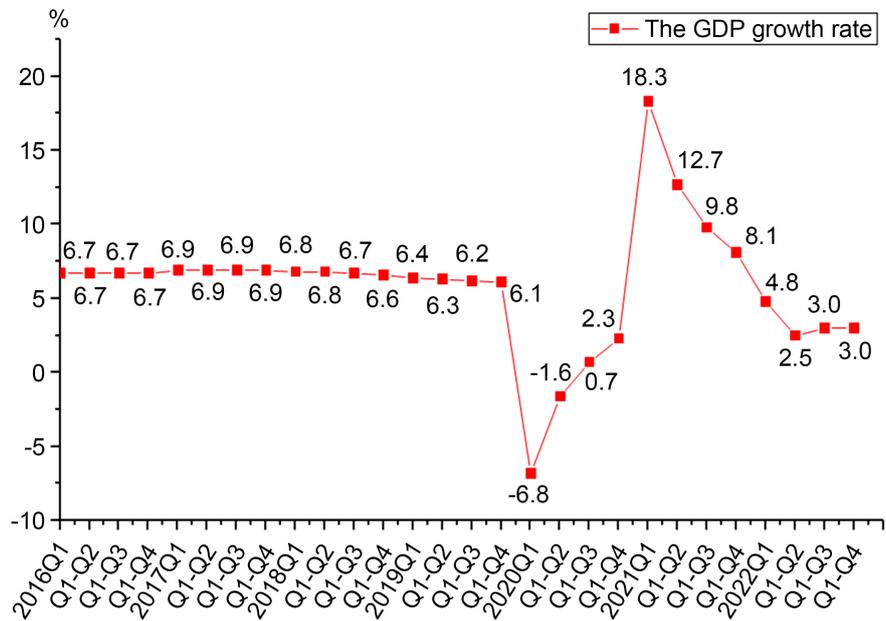

Figure 4. The GDP growth rate. (created by authors).

The consumption upgrade and guidance of the higher and middle class is not a difficult task that needs special discussion (tax reform is a necessary means to achieve a more reasonable redistribution of wealth and to guide the reasonable flow of capital), but the most difficult is the consumption upgrade of the vast proportion of the lower income group, especially the farmer group (Qiang & Hu, 2022).

The current domestic mainstream fund managers are positioning the main force of consumption on the new middle class group, but the team in this paper does not fully agree. First of all, the main group of consumption before the pandemic was mainly the middle and higher class in the amount of 100 - 300 million (Zhou et al., 2022), geographically skewed to the southeast coastal cities and the four regions of Beijing, Tianjin, Hebei, Jiangsu, Zhejiang and Shanghai, Pearl River Delta and Sichuan and Chongqing. This group's consumption shrinkage in the pandemic years is mainly in the high-end consumption area, and combined with the more fragile balance sheet and mortgage pressure in first- and second-tier cities. It can be expected that this group's consumption recovery will be fast, but not strong enough, and we cannot even be blindly optimistic that they will create a stronger consumption volume and scale than before the pandemic. We believe that in the post-pandemic era, the main force of consumption should look more at the low-production group, especially the vast number of farmers. According to the common sense of economics, to enhance a person's consumption, he/she should first be given more income, then to enhance confidence in future income and stimulate his/her potential consumption demand (Shi & Jiao, 2022), and finally to be able to fully mobilize the transfer of personal precautionary savings to consumption cash to achieve the completion of consumption action.

On the one hand, China's vast base of farmers has a very high potential demand for consumption upgrade. With the improvement of material and cultural level since the reform and opening up (Su et al., 2023), the vast number of farmers' demand for material goods has been upgraded from necessary taste-type living goods to high-level living goods, especially the demand for high-standard self-built houses in rural areas, the demand for cars as a means of transportation in the county, the demand for new household appliances to improve living, the financial demand to improve education and medical level, and the demand for agricultural products based on the supply and marketing network is constantly rising, and this part of the demand is incomparable to other consumer groups in terms of quantity and variety, and is the newborn consumer army (Steinbock, 2012) (see Figure 5).

It can be said that the high consumption group will have to bear more social and tax responsibilities in the post-pandemic era, the middle class will complete the task of rapid recovery of consumption (that is, return to the past level), and the task of breaking through to new highs in consumption power can only be completed by the low production group under the new policy.

Our government's continuous layout of the new rural supply and marketing network after the 14th Five-Year Plan is based on the integration of resources from the national level and provincial cross-domain level, which will supply and sell the whole chain to close the loop and make up for the shortcomings of the new private model of the cyberstar economy and the internet economy for poverty alleviation, so that the work of helping the majority of farmers to preserve income promotion is more refined. Since the reform and opening up (Zhao et al., 2022), from the very beginning of the household responsibility system to the government's abolition of the arable land tax and breeding income preservation subsidies to help farmers, to the establishment of the agricultural futures market, to the promotion of inclusive finance and credit for the agriculture, to the recent construction of the new supply and marketing agency network, our government's efforts and breadth of work continue to deepen, with a clear goal to make the farmers' groups generally rich, so that their consumption power continues to rise, so that they are truly willing and assured. The government's goal is to make farmers more affluent, to increase their spending power, and to make them willing and comfortable to spend their precautionary savings to improve their lives (farmers are the group with the most precautionary savings).

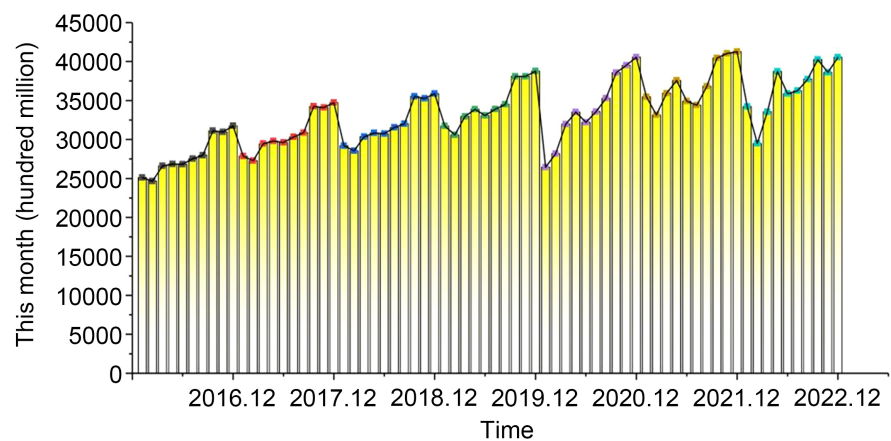

Figure 5. Total retail sales of social consumer goods. (created by authors).

Our research suggests that the Chinese government will stimulate the consumption of farmers by increasing both income (or transfer payments) and expenditure. In terms of increasing income (or transfer payments), in addition to preserving the price of agricultural products through the new supply and marketing agency network (Andolfatto, 2018) and the private agricultural economy model, it will also vigorously use large amounts and large percentages of consumption vouchers (e.g., promotional coupons), most likely in the form of digital RMB for precise and universal distribution (Auer et al., 2020), which can overlap to stimulate the development of China's digital economy (Solberg Söilen & Benhayoun, 2021); in terms of increasing spending, it will rely on farmers' major high consumption demand (Infante et al., 2022). In terms of increased spending, we will rely on the main high consumption needs of farmers, and cooperate with the majority of high-quality supply-side enterprises and platforms (2023 is also a promising year for the platform economy to bottom out), or use the digital RMB approach for precise poverty alleviation-style consumption matching.

In total, the country will increase higher responsibility for high-income groups from a tax perspective, elevate more promotional fee-preserving confidence initiatives for the middle class, cash in more income enhancement measures for low-income groups, especially rural groups with significant bases, and lead a new economic model that enhances their quality of life-a digital consumer model aligned with urban consumption group. The digital consumption index established in this paper can be used to measure and track the potential and function of digital consumption in the consumer recovery wave (Kiff et al., 2020).

## 4. Policies

Based on the above concept and logic, we propose the following policy recommendations.

We hope that in the first quarter of 2023, the central bank will cut interest rates or required reserve ratio (RRR) at least once more for the purpose of stimulating the economy, and that there will be a salary increase in the national civil service and institutional system (minimum income standards and wages in the state and private sectors will be raised simultaneously), and that there will be greater and more actions in the digital RMB overlaid with tax reform and consumption vouchers (Chiu et al., 2019), and that we may even explore channels and precise support for low-income people through digital fiat money (Ozili, 2021). We may even explore channels and solutions for precise support of low-income groups through digital fiat money (more social forces and solutions will emerge) (Davoodalhosseini, 2018).

The study of social finance, social consumption and total money supply M2 shows that under the main theme of consumer recovery, the CPI-linked consumer sector will be the first to rise in the first quarter, especially agricultural companies such as meat, fruits and vegetables will be the first to gain. Whether the CPI index can rise quickly or to what extent depends largely on the government's tolerance attitude. We believe an appropriate tolerance for a slower broad-based climb in CPI in the post-pandemic era will help boost the slope of China's economic bottoming out, especially based on the premise of lower CPI and GDP growth rates in 2022. The tasting-based and integrated services sectors rebounded faster during this period.

From the second quarter onwards, non-CPI closely linked consumer companies will benefit significantly from their results or report large positive earnings, especially the touching-based and hearing-based equipment sectors will benefit more.

From the third quarter to the fourth quarter, the tone of consumer recovery may shift to the beverage and alcoholic beverages category and the light luxury sense of smelling-based products. The slope of the sustained rise in this segment is not among the largest, but the time is longer and the space is more substantial.

The proportion of basic tasting-based consumption will decline throughout the year, the enjoyable taste consumption, especially alcoholic beverages will continue to rise, smelling-based consumption will appear faster up, superimposed on a variety of sensory audio-visual touch of digital consumption will continue to complete the task of recovery and innovation, and comprehensive service consumption will have different degrees of recovery. From the point of view of duration and upside, digital consumption is the main force to be reckoned with.

## 5. Summary

In fact, we can never accurately predict the trend of financial asset prices (He et al., 2021), but according to historical experience, there is a certain pattern of price fluctuations, especially in the process of recovery back to reasonable valuations.

Assuming that a period of time A-share field trading and off-site sideline funds are relatively fixed, the overall volume of these funds is firstly not enough to instantly and simultaneously make all reasonably valued stocks rise in tandem and keep the rate of rise consistent; secondly, the holders of funds (or decision makers) are divided into institutions, lobbyists, retail investors and the more mysterious national team, each of whom represents long term style funds, short term style funds, emotional style funds and regulator's will funds. Different styles of decision making are not possible to synchronize and consistent. Therefore, we can say that even if a sector or several sectors (including horizontal and vertical) in the A-share market is unanimously bullish in the middle and distal markets, there will be a near-end asynchronous rotation effect, and the supe-rimposed emotional factors of lobbyists and retail funds will enhance the magnitude of this rotation, presenting a clear trend rule between sectors in time sequence.

The research team in this paper will define this phenomenon as the Roller Conduction Effect[1], intuitive feeling like a bucket of water in the rolling rotation, because of tension and gravity, clutch force and other forces together, resulting in water kinetic energy in accordance with a direction of the overall rolling conduction, in a particular point on the cylinder wall is to see the continuous cycle of counter-balance.

Investors, especially the majority of retail investors without capital advantages and information advantages, in the non-exhaustive market are not to chase the roller in the water kinetic energy-excessive pursuit of frequent trading state of chasing up and down, the most sensible approach is to patiently hold and increase positions in a certain point asset, waiting for the roller conduction to this point to achieve profits.

Finally, we sincerely hope that the economic development of the motherland can achieve a high-quality recovery, the early realization of Chinese modernization!

## Conflicts of Interest

The authors declare no conflicts of interest regarding the publication of this paper.

## References

Agur, I., Ari, A., & Dell'Ariccia, G. (2019). *Designing Central Bank Digital Currencies*. CEPR Discussion Paper Series. https://doi.org/10.5089/9781513519883.001

Andolfatto, D. (2018). Assessing the Impact of Central Bank Digital Currency on Private Banks. *The Economic Journal, 131,* 525-540. https://doi.org/10.20955/wp.2018.026

Auer, R. A., Cornelli, G., & Frost, J. (2020). *Rise of the Central Bank Digital Currencies: Drivers, Approaches and Technologies*. CEPR Discussion Paper Series. https://doi.org/10.2139/ssrn.3724070

Boateng, A., Nguyen, V. H. T., Du, M., & Kwabi, F. O. (2022). The Impact of CEO Compensation and Excess Reserves on Bank Risk-Taking: The Moderating Role of Monetary Policy. *Empirical Economics, 62,* 1575-1598. https://doi.org/10.1007/s00181-021-02086-4

Cai, F., & Zhang, X. (2022). Following the Most Sensible and Logical Course of the New Normal of Economic Development. In F. Cai, & X. J. Zhang (Eds.), *Constructing Political Economy with Chinese Characteristics* (pp 87-115). Springer. https://doi.org/10.1007/978-981-19-2824-6_3

Calinescu, T., Likhonosova, G., & Zelenko, O. (2023). Circular Economy: Ukraine's Reserves and the Consequences of the Global Recession. In V. Koval, Y. Kazancoglu, &

---

[1]The roller conduction effect is suitable for a period of time stock capital game prediction, if the breakthrough of this time limit, if there is a huge amount of incremental capital influx or withdrawal, such as the southward flow of funds to the A-share market action, will break through the cylinder wall, resulting in changes in the roller conduction effect.

E.-S. Lakatos (Eds.), *International Conference on Sustainable, Circular Management and Environmental Engineering* (pp. 238-251). Springer. https://doi.org/10.1007/978-3-031-23463-7_16

Castro, C. G., Trevisan, A. H., Pigosso, D. A., & Mascarenhas, J. (2022). The Rebound Effect of Circular Economy: Definitions, Mechanisms and a Research Agenda. *Journal of Cleaner Production, 345,* Article ID: 131136. https://doi.org/10.1016/j.jclepro.2022.131136

Chiu, J., Davoodalhosseini, S. M., Hua, J., & Zhu, Y. (2019). Central Bank Digital Currency and Banking. *SSRN Electronic Journal.* https://doi.org/10.2139/ssrn.3331135

Davoodalhosseini, S. M. (2018). *Central Bank Digital Currency and Monetary Policy*. PSN: Central Banks & Reserves (Topic). https://doi.org/10.2139/ssrn.3011401

Ding, C., Liu, C., Zheng, C., & Li, F. (2022). Digital Economy, Technological Innovation and High-Quality Economic Development: Based on Spatial Effect and Mediation Effect. *Sustainability, 14,* Article No. 216. https://doi.org/10.3390/su14010216

Hall, S. G., Tavlas, G. S., & Wang, Y. (2023). Drivers and Spillover Effects of Inflation: The United States, the Euro Area, and the United Kingdom. *Journal of International Money and Finance, 131,* Article ID: 102776. https://doi.org/10.1016/j.jimonfin.2022.102776

He, C., Chen, R., Xue, B., & He, M. (2021). Investor Sentiment, Limited Arbitrage and Stock Price Anomalies. *Economic Research Journal, 1,* 58-73.

Infante, S., Kim, K., Orlik, A. B., Silva, A., & Tetlow, R. J. (2022). *The Macroeconomic Implications of CBDC: A Review of the Literature*. Finance and Economics Discussion Series.

Kahn, C. M., Oordt, M. R., & Zhu, Y. (2021). Best Before? Expiring Central Bank Digital Currency and Loss Recovery (Extended Abstract). In *International Conference on Blockchain Economics, Security and Protocols*.

Kiff, J.S., Alwazir, J., Davidovic, S., Farias, A., Khan, A., Khiaonarong, T., Malaika, M., Monroe, H., Sugimoto, N., Tourpe, H., & Zhou, P. (2020). *A Survey of Research on Retail Central Bank Digital Currency*. IMF Working Paper WP/20/104. https://doi.org/10.2139/ssrn.3652492

Kuckertz, A., Brändle, L., Gaudig, A., Hinderer, S., Reyes, C. A. M., Prochotta, A., Steinbrink, K. M., & Berger, E. S. (2020). Startups in Times of Crisis—A Rapid Response to the COVID-19 Pandemic. *Journal of Business Venturing Insights, 13,* e00169. https://doi.org/10.1016/j.jbvi.2020.e00169

Langdana, F. K. (2022). Budget Deficits, Trade Deficits, and Global Capital Flows: The National Savings Identity. In F. K. Langdana (Ed.), *Macroeconomic Policy* (pp. 27-56). Springer. https://doi.org/10.1007/978-3-030-92058-6_3

Liu, Z., & Liu, X. (2022). Is China's Infrastructure Development Experience Unique? *Journal of Chinese Economic and Business Studies,* 1-18. https://doi.org/10.1080/14765284.2022.2040074

Lyu, W., & Chen, X. (2023). *Reshaping Singapore's Tourism Industry after the Pandemic*. https://doi.org/10.2139/ssrn.4352531

Meaning, J., Dyson, B., Barker, J., & Clayton, E. (2018). *Broadening Narrow Money: Monetary Policy with a Central Bank Digital Currency*. Bank of England Working Paper No. 724. https://doi.org/10.2139/ssrn.3180720

Minesso, M. F., Mehl, A., & Stracca, L. (2020). Central Bank Digital Currency in an Open Economy. *Journal of Monetary Economics, 127,* 54-68. https://doi.org/10.2139/ssrn.3733463

Nicola, M., Alsafi, Z., Sohrabi, C., Kerwan, A., Al-Jabir, A., Iosifidis, C., Agha, M., & Agha, R. (2020). The Socio-Economic Implications of the Coronavirus Pandemic (COVID-19): A Review. *International Journal of Surgery, 78,* 185-193. https://doi.org/10.1016/j.ijsu.2020.04.018

Ozili, P. K. (2021). Central Bank Digital Currency in Nigeria: Opportunities and Risks. In S. Grima, E. Özen, & H. Boz (Eds.), *The New Digital Era: Digitalisation, Emerging Risks and Opportunities (Contemporary Studies in Economic and Financial Analysis, Vol. 109A)* (pp. 125-133). Emerald Publishing Limited. https://doi.org/10.2139/ssrn.3917936

Pieterse, J. N. (2002). Global Inequality: Bringing Politics Back in. *Third World Quarterly, 23,* 1023-1046. https://doi.org/10.1080/0143659022000036667

Prasad, E. S. (2021). *The Case for Central Bank Digital Currencies*.

Qiang, H., & Hu, L. (2022). Population and Capital Flows in Metropolitan Beijing, China: Empirical Evidence from the Past 30 Years. *Cities, 120,* Article ID: 103464. https://doi.org/10.1016/j.cities.2021.103464

Shi, H., & Jiao, Z. (2022). The Research on the Relationship between Consumption Downturn and COVID-19—Analysis of Survey Data from China. *International Journal of Business and Management, 17,* 1-90. https://doi.org/10.5539/ijbm.v17n12p90

Solberg Söilen, K., & Benhayoun, L. (2021). Household Acceptance of Central Bank Digital Currency: The Role of Institutional Trust. *International Journal of Bank Marketing, 40,* 172-196. https://doi.org/10.1108/IJBM-04-2021-0156

Son, J., Bilgin, M. H., & Ryu, D. (2022). Consumer Choices under New Payment Methods. *Financial Innovation, 8,* Article No. 82. https://doi.org/10.1186/s40854-022-00387-w

Steinbock, D. (2012). The Eurozone Debt Crisis: Prospects for Europe, China, and the United States. *American Foreign Policy Interests, 34,* 34-42. https://doi.org/10.1080/10803920.2012.653719

Su, C., Lyu, W., & Chen, X. (2023). Macroeconomic Forecast and Investment Decisions for China in 2023. *Research Ambition, 7*.

Su, C., Lyu, W., & Liu, Y. (2022). *The Relationship between Digital RMB and Digital Economy in China*. https://doi.org/10.2139/ssrn.4208078

Zhao, S., Peng, D., Wen, H., & Wu, Y. (2022). Nonlinear and Spatial Spillover Effects of the Digital Economy on Green Total Factor Energy Efficiency: Evidence from 281 Cities in China. *Environmental Science and Pollution Research,* 1-21. https://doi.org/10.1007/s11356-022-22694-6

Zhou, C., Gong, M., Xu, Z., & Qu, S. (2022). Urban Scaling Patterns for Sustainable Development Goals Related to Water, Energy, Infrastructure, and Society in China. *Resources, Conservation and Recycling, 185,* Article ID: 106443. https://doi.org/10.1016/j.resconrec.2022.106443